\documentclass[fleqn,twoside]{article}
\usepackage{espcrc2}

\usepackage{graphicx}
\usepackage{amsfonts, color, float, bbm, amsmath}
\usepackage{amssymb,amsthm}

\title{A critical look at $V_{us}$ determinations
from hadronic $\tau$ decay data}

\author{Kim Maltman, Math and Statistics, York University,
        4700 Keele St., Toronto, ON CANADA M3J 1P3%
        \thanks{Alternate address: CSSM, Univ. of Adelaide, Adelaide, SA
                5005 Australia; work supported by the Natural Sciences
                and Engineering Research Council of Canada}}

\begin{document}

\begin{abstract}
A critical review of hadronic $\tau$ decay data based determinations 
of $\vert V_{us}\vert$ is given, focussing 
on the impact of the slow convergence of the integrated $D=2$ OPE series for
the conventional flavor-breaking sum rule determination and 
the potential role of as-yet-unmeasured multiparticle contributions 
to the strange spectral distribution. Additional information obtainable
from analyses of inclusive strange decay data alone, and from mixed
electroproduction-$\tau$ sum rules with much reduced OPE uncertainties,
is also discussed. Self-consistency tests are shown to favor
determinations which reduce somewhat discrepancies with 3-family
unitarity expectations.
\vspace{1pc}
\end{abstract}

\maketitle

\section{Introduction}

$K_{\ell 3}$ and $\Gamma [K\rightarrow\mu\nu ]/\Gamma [\pi\rightarrow\mu\nu ]$
based determinations of $\vert V_{us}\vert$~\cite{kell3andKratiosvus} 
are in excellent agreement with expectations
based on $\vert V_{ud}\vert$~\cite{htvud} and 3-family
unitarity. In contrast,
determinations from flavor breaking (FB) sum rules involving
hadronic $\tau$ decay data~\cite{gamizetal,kmcwvus,gamizetalnew,hfag10} 
yield results $\sim 3\sigma$ low, even after 
accounting for recent HFAG strange exclusive branching fraction (BF)
updates~\cite{hfag10}.

For a kinematic-singularity-free correlator, $\Pi$, with spectral 
function, $\rho (s)$, the finite energy sum rule (FESR) relation on which
the FB $\tau$ determination is based takes the form
\begin{equation}
\int_0^{s_0}w(s) \rho(s)\, ds\, =\, -{\frac{1}{2\pi i}}\oint_{\vert
s\vert =s_0}w(s) \Pi (s)\, ds\ \ \ 
\label{basicfesr}
\end{equation}
a result valid for any $s_0$ and any analytic $w(s)$.

$\vert V_{us}\vert$ is obtained by applying Eq.~(\ref{basicfesr}) to
the FB difference $\Delta\Pi_\tau \, \equiv\,
\left[ \Pi_{V+A;ud}^{(0+1)}\, -\, \Pi_{V+A;us}^{(0+1)}\right]$,
where $\Pi^{(J)}_{V/A;ij}(s)$ are the spin $J=0,1$ components
of the flavor $ij$, vector (V) or axial vector (A) current two-point
functions, and $(0+1)$ denotes the sum of $J=0$ and $1$ components.
For sufficiently large $s_0$, the RHS of Eq.~(\ref{basicfesr}) is 
evaluated using the OPE representation, $\left[\Delta\Pi_\tau\right]^{OPE}$,
while, for $s_0\leq m_\tau^2$, the LHS is obtainable from inclusive hadronic 
$\tau$ decay distributions. Explicitly, the spectral functions, 
$\rho^{(J)}_{V/A;ij}$, are related to the differential distributions, 
$dR_{V/A;ij}/ds$, of the normalized flavor $ij$ V or A current 
induced decay widths,
$R_{V/A;ij}\, \equiv\, \Gamma [\tau^- \rightarrow \nu_\tau
\, {\rm hadrons}_{V/A;ij}\, (\gamma )]/ \Gamma [\tau^- \rightarrow
\nu_\tau e^- {\bar \nu}_e (\gamma)]$, by~\cite{tsai}
\begin{eqnarray}
&&{\frac{dR_{V/A;ij}}{ds}}\, =\, c^{EW}_\tau \vert V_{ij}\vert^2
\left[ w_\tau (y_\tau ) \rho_{V/A;ij}^{(0+1)}(s)\ \ 
\right.\nonumber\\
&&\left. \qquad\qquad\qquad 
- w_L (y_\tau )\rho_{V/A;ij}^{(0)}(s) \right]
\label{basictaudecay}\end{eqnarray}
with $y_\tau =s/m_\tau^2$, $w_\tau (y)=(1-y)^2(1+2y)$,
$w_L(y)=2y(1-y)^2$, $V_{ij}$ the flavor $ij$ CKM matrix element,
and, with $S_{EW}$ a short-distance electroweak correction~\cite{erler02},
$c^{EW}_\tau \equiv 12\pi^2 S_{EW}/m_\tau^2$.

The choice of FESRs involving the $J=0+1$ combination, $\Delta\Pi_\tau$, 
rather than that corresponding to the linear combination of 
spectral functions appearing in  Eq.~(\ref{basictaudecay}),
is predicated on the extremely bad behavior of the integrated
$J=0$, $D=2$ OPE series for scales kinematically
accessible in $\tau$ decay~\cite{longprob}.
Fortunately, the dominant $J=0$ spectral contributions
are the accurately known, non-chirally-suppressed 
$\pi$ and $K$ pole terms. The remaining, continuum
contributions, are (i) doubly chirally suppressed ($\rho_{V+A;ij}^{(0)}
\propto [(m_i\mp m_j)^2]$), (ii) numerically negligible for $ij=ud$, 
and (iii) determinable phenomenologically for
$ij=us$ $J=0$ via dispersive~\cite{jop} and sum rule~\cite{mksps} analyses 
of the strange scalar and pseudoscalar channels. The $J=0$
contributions can thus be subtracted bin by bin from $dR_{V+A;ij}/ds$,
allowing one to determine $\rho_{V+A:ij}^{(0+1)}(s)$
and construct the re-weighted $J=0+1$ spectral integrals,
$R^w_{V+A;ij}(s_0)$, and FB differences, $\delta R^w_{V+A}(s_0)$,
defined by
\begin{eqnarray}
&&{\frac{R^w_{V+A;ij}(s_0)}{c^{EW}_\tau \vert V_{ij}\vert^2}}\,
\equiv\, \int_0^{s_0}ds\, w(s)\, \rho^{(0+1)}_{V+A;ij}(s)\ ,\nonumber\\
&&\delta R^w_{V+A}(s_0)\, =\,
{\frac{R^w_{V+A;ud}(s_0)}{\vert V_{ud}\vert^2}}
\, -\, {\frac{R^w_{V+A;us}(s_0)}{\vert V_{us}\vert^2}}\nonumber\\
&&\qquad\qquad =\,c^{EW}_\tau\, \int_0^{s_0}ds\, w(s)\Delta\rho_\tau (s)\ .
\end{eqnarray}
Taking $\vert V_{ud}\vert$ and any parameters in the OPE
representation of $\delta R_{V+A}^{w}$ from other sources, 
Eq.~(\ref{basicfesr}) then yields~\cite{gamizetal}
\begin{equation}
\vert V_{us}\vert \, =\, \sqrt{{\frac{R^w_{V+A;us}(s_0)}
{{\frac{R^w_{V+A;ud}(s_0)}{\vert V_{ud}\vert^2}}
\, -\, \delta R^{w,OPE}_{V+A}(s_0)}}}\ .
\label{tauvussolution}\end{equation}
For weights used previously in the
literature~\cite{gamizetal,kmcwvus,gamizetalnew}, the OPE
contribution to the denominator 
is at the few-to-several-$\%$ level of the $ud$ spectral
integral term, making modest accuracy for $\delta R^{w,OPE}_{V+A}(s_0)$
sufficient for a high accuracy determination of $\vert V_{us}\vert$.
The smallness of the FB OPE corrections is illustrated
by the results of the $s_0=m_\tau^2$, $w_\tau$ FESR where,
with updated HFAG $\tau$ BFs~\cite{hfag10}, supplemented
by Standard Model (SM) $K_{\mu 2}$ expectations for 
the $\tau\rightarrow K\nu_\tau$ BF, $B_K$, 
the value $\vert V_{us}\vert =0.2108(19)$ obtained ignoring 
FB OPE corrections differs from that obtained using one of the 
possible evaluations of $\delta R^{w_\tau,OPE}_{V+A}(m_\tau^2)$,
$\vert V_{us}\vert =0.2174(22)$~\cite{hfag10}, by only $\sim 3\%$.


The task of quantifying the uncertainty in the OPE estimate for
$\delta R^{w_\tau}_{V+A}(s_0)$ and, from this, the theoretical error 
on $\vert V_{us}\vert$, is complicated by the rather slow convergence, 
at the correlator level, of the series for the leading dimension $D=2$ 
OPE contribution $\left[\Delta \Pi_\tau \right]_{D=2}^{OPE}$. 
To four loops~\cite{bckd2ope}
\begin{eqnarray}
&&\left[\Delta\Pi_\tau (Q^2)\right]^{OPE}_{D=2}\, =\, {\frac{3}{2\pi^2}}\,
{\frac{m_s(Q^2)}{Q^2}} \left[ 1\, +\, {\frac{7}{3}} \bar{a}\ \ \ 
 \right.\nonumber
\\
&&\left. \ \  +\, 19.93 \bar{a}^2 \, +\, 208.75 \bar{a}^3
\, +\, d_4 \bar{a}^4\, +\, \cdots \right]\ \ \ 
\label{d2form}\end{eqnarray}
with $\bar{a}=\alpha_s(Q^2)/\pi$, and $\alpha_s(Q^2)$ and $m_s(Q^2)$
the running coupling and strange quark mass in the $\overline{MS}$ scheme
{\footnote{In what follows, we employ the estimate 
$d_4\sim 2378$~\cite{bckd2ope} for the as-yet-undetermined
5-loop coefficient $d_4$.}}.
Since $\bar{a}(m_\tau^2)\simeq 0.1$, convergence at the spacelike point on
the contour $\vert s\vert = s_0$ is marginal at best. With such slow 
convergence, conventional error estimates may significantly 
underestimate the $D=2$ truncation uncertainty. 

Fortunately, the FESR framework allows for internal self-consistency checks.
Assuming both the data and OPE error estimates are reliable, the
$\vert V_{us}\vert$ obtained from Eq.~(\ref{tauvussolution}) should 
be independent of $s_0$ and $w(s)$. On the 
OPE side, three common methods can be employed for evaluating 
the integrated $D=2$ contribution: the contour improved (CIPT) 
prescription, used with either (i) the truncated expression for
$\left[ \Delta\Pi_\tau\right]^{OPE}_{D=2}$ itself, or, (ii) after
partial integration, with the correspondingly truncated Adler function,
and (iii) the truncated fixed-order (FOPT) prescription
(for which the correlator and Adler function results are identical).
While the CIPT prescription corresponds to the conventional procedure
of choosing a local scale, arguments in favor of the FOPT alternative
also exist~\cite{bj08}. At a given truncation order, the
three prescriptions differ only by contributions of yet higher order. If
the estimated truncation errors are reliable, the results
for $\vert V_{us}\vert$ obtained using the different prescriptions 
should thus also agree within these errors.

\section{$\vert V_{us}\vert$ from FB hadronic $\tau$ decay FESRs}

The results in this section are obtained using the updated December 2010
HFAG $\tau$ BFs~\cite{hfag10}, supplemented by SM $K_{\mu 2}$
expectations for $B_K$. SM $\pi_{\mu 2}$ expectations
are similarly employed for the non-strange $\pi$ pole contributions.
Assuming lepton universality, the constrained HFAG determination
of the electron BF, $B_e=0.17852(27)$, implies
$R_{V+A;ud}=3.467(9)$ and $R_{V+A;us}=0.1623(28)$. 
The change in $R_{V+A;ud}$ necessitates a small rescaling of
the publicly available 2005 ALEPH $ud$ distributions~\cite{alephud05}, 
from which the continuum contributions to $\rho_{V/A;ud}(s)$ are obtained. 
For the continuum part of $\rho_{V+A;us}$, the BaBar and Belle 
analyses of the inclusive strange distribution are not yet complete. The 
completed 1999 ALEPH distribution~\cite{alephus99}
corresponds to exclusive strange BFs with significantly larger errors,
and, in many cases, significantly different central values, than 
those obtained by the B factory experiments~\cite{hfag10}. 
As an interim measure, we follow the strategy of 
Ref.~\cite{alephusrescaleidea}, ``partially updating'' $\rho_{V+A;us}(s)$ 
to reflect the new values of the BFs through a mode-by-mode rescaling 
of the 1999 ALEPH distribution. This proceedure, though not ideal, 
has been tested using the BaBar distribution data for  
$\tau\rightarrow K^-\pi^+\pi^-\nu_\tau$~\cite{babarkppallchg}
(a mode having a particularly large BF change), and found to yield spectral 
integral contributions in surprisingly good agreement with those of the 
actual BaBar data. The rescaling method, however, has not been
checked for other modes, and is likely to be less reliable for the
$K3\pi ,\, K4\pi ,\cdots$ contributions,
where the ALEPH distribution was not measured, but
estimated using Monte Carlo, and has a purely phase space shape.

OPE input is as in the last of Refs.~\cite{kmcwvus}, except for the 
update $\alpha_s^{n_f=3}(m_\tau^2)=0.3181(57)$, which reflects the
new world average, $\alpha_s^{n_f=5}(M_Z^2)=0.1184(7)$~\cite{bethke09}.

We begin with the conventional $w_\tau$ determination. For
$s_0=m_\tau^2$, the $ud$ and $us$ spectral integrals are determined 
by the corresponding inclusive BFs. Improvements 
to the various exclusive strange BFs~\cite{hfag10} then 
translate into an improved determination of the $us$ spectral integral, even
without completion of the remeasurement of the inclusive strange distribution.
This is {\it not} the case for other $s_0$ and/or other $w(s)$. 
The $D=2$ truncation uncertainty in this case has been estimated
using standard last-term-retained$\oplus$residual-scale-dependence methods
(known to work well for perturbative series displaying good 
convergence behavior). The resulting combined theoretical uncertainty on 
$\vert V_{us}\vert$ is then $0.0005$~\cite{gamizetalnew}. 
To test for the independence of $\vert V_{us}\vert$ on $s_0$ and $w(s)$,
one must go beyond $w_\tau$ and/or $s_0=m_\tau^2$. 
We perform these tests using the interim updated $\rho_{V+A;us}(s)$.

\begin{figure}[!ht]
\rotatebox{270}{\mbox{
\includegraphics*[width=1\columnwidth,angle=0]
{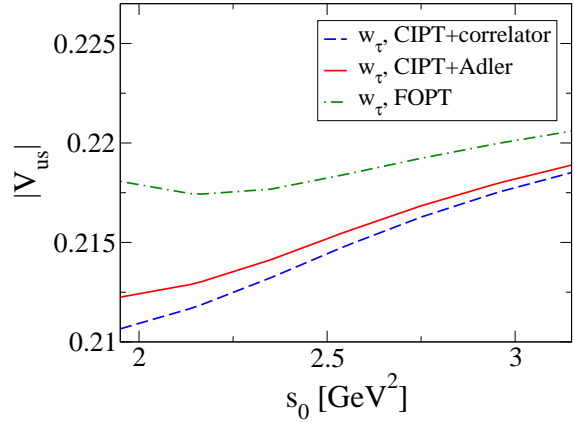}
}}
\vspace{-1.2cm}
\caption{$\vert V_{us}\vert$ vs. $s_0$ from the FB $w_\tau$-weighted 
FESR for the three different prescriptions for handling the integrated
$D=2$ OPE series. }
\label{wtauvusfig}
\end{figure}

In Fig.~\ref{wtauvusfig}, results for $\vert V_{us}\vert$ 
as a function of $s_0$ are shown for each of the three
prescriptions for the $w_\tau$-weighted $D=2$ OPE series.
The two CIPT-based prescriptions give similar results, 
but show poor $s_0$-stability. The FOPT prescription
yields significantly improved, though not perfect, $s_0$-stability. 
For all $s_0$ the FOPT and CIPT results 
differ by significantly more than the $0.0005$ total theoretical 
error estimate mentioned above.

The observed $s_0$-instability of the FB $w_\tau$
FESR results could arise from deficiencies in any
of the contributions appearing in Eq.~(\ref{tauvussolution}).
Problems with the $ud$ spectral integrals seem unlikely,
but can be tested for using FESRs for the $us$ V+A channel alone.
A problem with missing higher multiplicity $us$ spectral strength could also
be exposed by such an analysis, assuming the OPE representation
used to be reliable. Specifically, for $s_0$ large enough to include
some portion of the region of missing strength, $\vert V_{us}\vert$ 
would be too low, while for $s_0$ below the region of the
missing strength, $\vert V_{us}\vert$ should rise back to its
true value. The $us$ V+A FESRs are considered in the next section.

The most obvious candidate for a source of $s_0$-instability 
in the OPE contribution is the potentially slowly
converging integrated $D=2$ series. For 
$s_0=m_\tau^2$, the $w_\tau$-weighted integrated $D=2$ series behaves as
$1+0.29+0.10-0.04-(0.20)+\cdots $ for the CIPT+Adler function
case, $1+0.15+0.02-0.12-(0.29)+\cdots$ for the CIPT+correlator
case, and $1+0.40+0.26+0.15+(0.08)+\cdots $ for the FOPT case,
where the terms in parentheses are the estimated 5-loop,
$O(\alpha_s^4)$ contributions. For the two CIPT
cases, though there is cancellation on the contour at intermediate
orders, this cancellation does not persist to higher orders,
and the integrated series appears rather badly behaved. The
behavior is better for the FOPT case, which was also found to yield
improved $s_0$-stability. Worth noting
is the fact that $\delta R^{w_\tau,OPE}_{V+A}(m_\tau^2)$ for
the better behaved FOPT prescription is 
a factor of $\sim 2$ larger than that for
either of the two CIPT prescriptions. The $\sim 100\%$ increase
between CIPT and FOPT increases $\vert V_{us}\vert$ by $\sim 0.0020$ 
establishing the insufficiently conservative nature of the
$0.0005$ estimate for the combined theoretical error. 
With the difference between the FOPT result 
and the average of the two CIPT results as a new truncation uncertainty
estimate, this source dominates the theoretical error.
Taking the FOPT result (favored by the $s_0$-stability criterion) as
the new central value, the result for the FB $w_\tau$ determination becomes
\begin{equation}
\vert V_{us}\vert = 0.2193(3)_{ud}(19)_{us}(19)_{th}\ ,
\label{fbtaufinalvus}
\end{equation}
$\sim 2.3\sigma$ below 3-family unitarity expectations.

\begin{figure}[!ht]
\rotatebox{270}{\mbox{
\includegraphics[width=1\columnwidth,angle=0]
{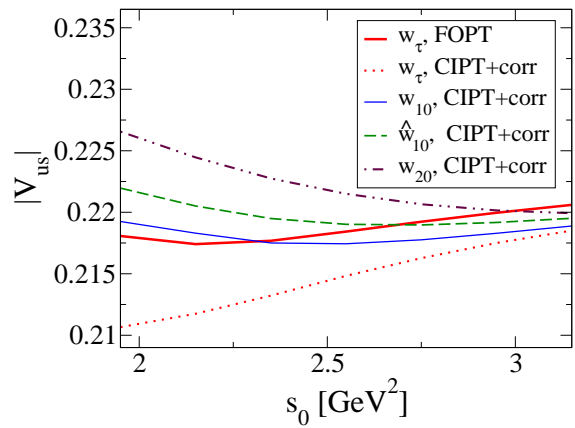}
}}
\vspace{-1.2cm}
\caption{$\vert V_{us}\vert$ vs. $s_0$, obtained using
the CIPT+correlator $D=2$ OPE prescription for
the FB $w_{10}$, $\hat{w}_{10}$ and $w_{20}$ FESRs, and, for comparison,
using FOPT and CIPT+correlator treatments for the $w_\tau$ FESR.}
\label{udusvusvariouswts}
\end{figure}

FB $\tau$-based FESRs can also be constructed using other $w(s)$.
For the CIPT prescription, if the
$s_0$-instability of the $w_\tau$ FESR results from
premature truncation of the $D=2$ series, convergence can be improved 
by choosing weights which emphasize contributions from
the region of the complex $s=-Q^2$ plane away from the spacelike
point, where $\vert \alpha_s(Q^2)\vert$ is smaller and the convergence,
at the correlator level, of the $D=2$ series better. Three weights (denoted
$w_{10}(y)$, $\hat{w}_{10}(y)$, and $w_{20}(y)$, with $y=s/s_0$)
having this, as well as other desirable properties, were constructed
in Ref.~\cite{km00}. The results for $\vert V_{us}\vert$ from the
resulting FB FESRs, using the CIPT+correlator prescription, are
shown in Fig.~\ref{udusvusvariouswts}. The $w_\tau$ results for both
the FOPT and CIPT+correlator prescriptions are shown for comparison.
Improved $s_0$-stability compared to the $w_\tau$ CIPT results is
observed{\footnote{FOPT-based results 
for the new weights are not shown, though they also display significantly 
improved $s_0$-stability. The reason is that,
having been constructed to improve the convergence
of the integrated CIPT series, there is no reason to
expect the new weights to also improve the convergence of the
integrated FOPT series, and indeed they do not. For example,
the FOPT version of the $\hat{w}_{10}$-weighted $D=2$ series 
behaves as $1+0.51+0.43+0.39+(0.41)+\cdots$, in contrast
to the CIPT+correlator version, which behaves as 
$1+0.24+0.19+0.15+(0.11)+\cdots$.}}. The $\vert V_{us}\vert$ values
are compatible with the result of Eq.~(\ref{fbtaufinalvus})
within the theoretical errors quoted previously for that case.
The central result for $\hat{w}_{10}$ (the weight showing the best 
$s_0$-stability) is $\vert V_{us}\vert = 0.2188$. The experimental
error, which, absent a re-measurement of the inclusive strange distribution,
has to be based on the 1999 ALEPH $us$ covariances, is $0.0033$.

\section{The $us$ V+A and mixed $\tau$-EM FESRs}

\begin{figure}[!ht]
\rotatebox{270}{\mbox{
\includegraphics[width=1\columnwidth,angle=0]
{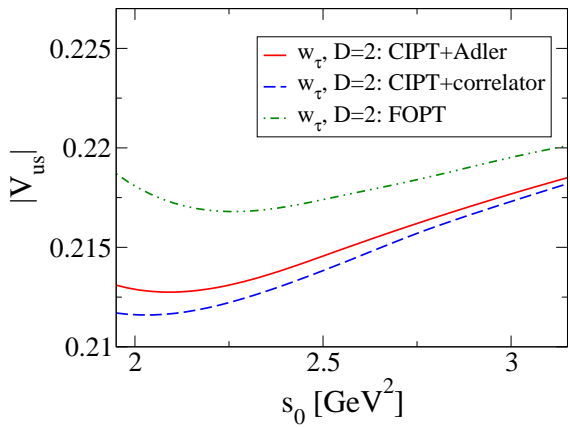}
}}
\vspace{-1.2cm}
\caption{$\vert V_{us}\vert$ vs. $s_0$ for the
$w_\tau$-based $us$ V+A FESR, using the three different prescriptions 
for handling the integrated $D=2$ OPE series. }
\label{usvpavusfig}
\end{figure}

FESR analyses of the $us$, $J=0+1$, V+A correlator combination
provide an alternate determination of $\vert V_{us}\vert$, one
that provides some cross-checks on the FB FESR results. Two new
ingredients enter the OPE side of these FESRs: $D=0$
contributions, determined by $\alpha_s$, and a $D=4$ gluon
condensate contribution absent from the FB difference $\Delta\Pi_\tau$.
It is known that, with the 5-loop $D=0$ expansion
as input~\cite{bckd0ope08}, pinched FESR analyses of the 
$ud$, $J=0+1$ V, A and V+A correlator combinations~\cite{my08} 
yield values of $\alpha_s$ in excellent agreement with current high-precison
lattice determinations~\cite{newlattice} and the new 
world average~\cite{bethke09}. A strong anti-correlation
between $\alpha_s$ and the gluon condenstate~\cite{my08} implies
$\langle \alpha_s G^2/\pi\rangle = 0.012\ GeV^4$ for the
central $\alpha_s$ input noted above. Results for $\vert V_{us}\vert$
as a function of $s_0$ obtained from the $w_\tau$-weighted 
FESR using each of three prescriptions for the integrated
$D=2$ series are presented in Fig.~\ref{usvpavusfig}. 
The compatibility of the results for each prescription with
those of the corresponding FB $w_\tau$
FESR is excellent. However, the $s_0$-dependence of $\vert V_{us}\vert$ 
for the two CIPT prescriptions is clearly not compatible with the assumption
that the $D=2$ OPE representation is reliable and the instability
in the FB $w_\tau$ FESR results for $\vert V_{us}\vert$ due 
solely to missing higher
multiplicity $us$ spectral strength. As for the FB
FESR case, the FOPT $D=2$ treatment produces improved, though not perfect, 
$s_0$-stability. 

With problems in the FB $\tau$ FESRs, to at
least some extent, due to slow $D=2$ OPE convergence,
FESRs with $D=2$ OPE contributions 
reduced at the correlator level are highly desirable. 
With this in mind, combinations of $\Pi^{(0+1)}_{V/A;ud}$, 
$\Pi^{(0+1)}_{V+A;us}$ and the EM correlator, $\Pi_{EM}$,
(whose spectral function, $\rho_{EM}$, is determined 
by the bare $e^+e^-\rightarrow hadrons$ cross-sections)
can be constructed having zero $D=0$ and vanishing 
$O(\alpha_s^0)$ $D=2$ OPE contributions~\cite{kmtauem08}. 
The unique such FB combination with the same
normalization for $\Pi^{(0+1)}_{V+A;us}$ as $\Delta\Pi_\tau$ is 
\begin{equation}
\Delta\Pi_M= 9\Pi_{EM} -6\Pi^{(0+1)}_{V;ud} + \Delta\Pi_\tau
\end{equation}
The $D=2$ OPE series for $\Delta\Pi_M(Q^2)$ is
\begin{equation}
{\frac{3}{2\pi^2}}\,
{\frac{\bar{m}_s}{Q^2}} \left[ {\frac{1}{3}} \bar{a}
+4.38\bar{a}^2  + 44.9 \bar{a}^3+\cdots\right]\ \ 
\label{tauemd2form}\end{equation}
which has not only (by construction) a vanishing $O(\alpha^0_s)$ coefficient, 
but also higher order coefficients significantly smaller than those
for $\Delta\Pi_\tau$, {\it c.f.} Eq.~(\ref{d2form}). The resulting 
strong suppression of the $D=2$ series
is accompanied by a fortuitous suppression of $D=4$ contributions.
Explicitly, with $\delta_4\equiv \langle m_s\bar{s}s
-m_\ell \bar{\ell}\ell\rangle$, the $D=4$ series for $\Delta\Pi_\tau$ 
and $\Delta\Pi_M$ have the form
${\frac{\delta_4}{Q^4}}\sum_k c_k\bar{a}^k$
with $(c_0, c_1, c_2)=(-2,-2,-26/3)$ for
$\Delta\Pi_\tau$ and $(0, 8/3, 59/3)$ for $\Delta\Pi_M$~\cite{bnp}.
Integrated $D>4$ contributions, which are expected to be
somewhat enhanced~\cite{kmtauem08}, can be fitted to data using their 
stronger $s_0$ dependences.

A complication for the $\Delta\Pi_M$ FESRs is the fact
that EM and $\tau$ results for the $2\pi$ and $4\pi$ components
of the $I=1$ V spectral function are not compatible at the level
of expected isospin-breaking corrections~\cite{newdavieretalgmuetc}.
We nonetheless present, as an illustration, the results for
$\vert V_{us}\vert$ as a function of $s_0$ for the $w_\tau$,
$w_2(y)=(1-y)^2$ and $w_3(y)=1-{\frac{3}{2}}y+{\frac{1}{2}}y^3$
$\Delta\Pi_M$ FESRs, obtained assuming the $\tau$ 
data to be correct in both cases. Because of the current
experimental complications, no attempt has been made to fit 
$D>4$ OPE contributions. Instead, $D=6$ contributions have
been estimated using the vacuum saturation approximation
and $D=8$ contributions neglected. The main 
point is to illustrate the significantly improved $s_0$-stability 
produced by the strong suppression of the integrated $D=2,4$ OPE 
contributions. The $s_0=m_\tau^2$, $w_\tau$ result in this case is
\begin{equation}
\vert V_{us}\vert = 0.2222(20)_{\tau}(28)_{EM}\, 
\end{equation}
where only the experimental errors have been displayed. 
While the very small theoretical errors for the mixed $\tau$-EM FESRs
make them promising for the future, further progress requires
a resolution of the experimental $2\pi$ and $4\pi$ problems.

\begin{figure}[!ht]
\rotatebox{270}{\mbox{
\includegraphics[width=1\columnwidth,angle=0]
{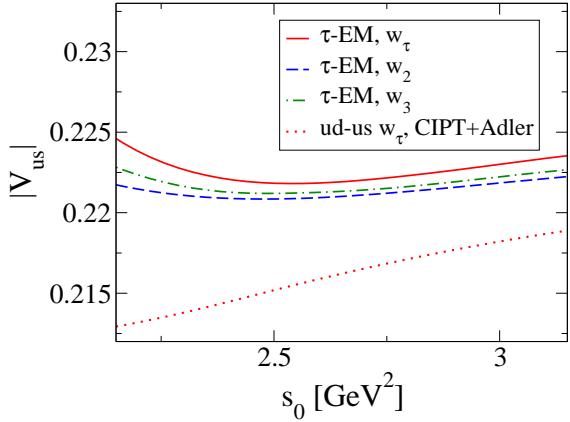}
}}
\vspace{-1.2cm}
\caption{$\vert V_{us}\vert$ vs. $s_0$ for
a selection of $\Delta\Pi_M$ FESRs and,
for comparison, the CIPT+Adler $w_\tau$ $\Delta\Pi_\tau$ 
determination.}
\label{emtaufig}
\end{figure}

\section{APPENDIX: Numerical results for the OPE input to 
the FB $s_0=m_\tau^2$, $w_\tau$ FESR}

This appendix contains numerical results for the OPE contributions
to the conventional inclusive, non-longitudinally-subtracted,
FB, $s_0=m_\tau^2$ FESR determination of $\vert V_{us}\vert$,
in a form that allows the reader to perform further explorations.
For this case, in the notation of Ref.~\cite{gamizetalnew}, one has
\begin{equation}
\vert V_{us}\vert \, =\, \sqrt{{\frac{R_{\tau ,S}}
{{\frac{R_{\tau ,V+A}}{\vert V_{ud}\vert^2}}
\, -\, \delta R_{\tau ,th}}}}\ ,
\label{appendixtauvussolution}\end{equation}
where $R_{\tau ,S}\equiv R_{V+A;us}$ and $R_{\tau ,V+A}\equiv
R_{V+A;ud}$ are the $B_e$-normalized inclusive strange and 
non-strange branching fractions, defined above
Eq.~(\ref{basictaudecay}). The theoretical contribution,
$\delta R_{\tau ,th}$, is a sum of $J=0$ (longitudinal) and $J=0+1\ (L+T)$ 
contributions, denoted $\delta R_\tau\vert^L$ and 
$\delta R_{\tau ,th}\vert^{L+T}$, respectively, in Ref.~\cite{gamizetalnew}.
The former, which must be determined phenomenologically because of
the problematic behavior of the longitudinal OPE representation,
is dominated by the accurately known $K$ and $\pi$ pole contributions.
Adding the current best assessment of the small continuum contributions
yields the result 
\begin{equation}
\delta R_\tau\vert^L = 0.1544\pm 0.0037
\label{longapp}\end{equation}
quoted in Ref.~\cite{gamizetalnew}.

$\delta R_{\tau ,th}\vert^{L+T}$ is evaluated using the OPE. The
OPE representation is a sum of contributions, $\delta R^{L+T}_{OPE;D}$, 
of dimensions $D=2,4,6\cdots$.
$D>6$ contributions are typically neglected in the literature. 
$\delta R^{L+T}_{OPE;D=2}$ is proportional to $m^2_s(2\ {\rm GeV})$ and, for 
$S_{EW}=1.0201$~\cite{erler02}, $m_\tau = 1.77677\pm 0.00015$
GeV~\cite{hfag10}, $m_s(2\ {\rm GeV})=94\ {\rm MeV}$, and the current 
$n_f=5$ world average $\alpha_s(M_Z^2)=0.1184\pm 0.007$~\cite{bethke09} 
(equivalent to the $n_f=3$ result $\alpha_s(m_\tau^2)=0.3181\pm 0.0057$)
has the values given in Table~\ref{appendixtable}, as a function of
the truncation order, $n_T$, and $D=2$ scheme choice. 

\begin{table}[!htb]
\caption{$\delta R^{L+T}_{OPE;D=2}$ for various $n_T$ and 
$D=2$ evaluation schemes 
(CI+Ad, CI+co and FOPT denote the CIPT+Adler function, 
CIPT+correlator and FOPT schemes, respectively)}
\label{appendixtable}
\newcommand{\m}{\hphantom{$-$}}
\newcommand{\cc}[1]{\multicolumn{1}{c}{#1}}
\renewcommand{\tabcolsep}{2pc} 
\renewcommand{\arraystretch}{1.2} 
\begin{tabular}{@{}lll}
\hline
Scheme&$n_T$&$\delta R^{L+T}_{OPE;D=2}$\\
\hline
CI+Ad&$3$&$0.0578\mp 0.0013$\\
&$4$&$0.0494\mp 0.0025$\\
CI+co&$3$&$0.0515\mp 0.0022$\\
&$4$&$0.0371\mp 0.0040$\\
FOPT&$3$&$0.1002\pm 0.0016$\\
&$4$&$0.1046\pm 0.0019$\\
\hline
\end{tabular}\\[2pt]
\end{table}

With ChPT input for
$m_s/(m_d+m_u)$, GMOR for $\langle (m_d+m_u)\bar{u}u\rangle$, and
the input already specified above, the $D=4$ contribution is given
by
\begin{eqnarray}
&&\delta R^{L+T}_{OPE;D=4}=0.0072(2)r_c-0.00030\qquad \nonumber\\
&&\qquad\quad 
+0.00060(11)\left({\frac{m_s(2\ {\rm GeV})}{94\ {\rm MeV}}}\right)^4\qquad 
\label{d4appendix}\end{eqnarray}
with $r_c\equiv \langle \bar{s}s\rangle /\langle \bar{u}u\rangle$,
and the three terms on the RHS being associated with the 
strange quark condensate, light quark condensate and $O(m_s^4)$
contributions, respectively. The errors shown are those
associated with the uncertainty in the input $\alpha_s$. The value 
$r_c=0.8$ employed in earlier analyses, follows from a 2002 sum rule 
analysis of $f_{B_s}/f_{B}$ with then-current quenched lattice results 
for the decay constant ratio as input~\cite{jaminlange}.
Current $n_f=2+1$ results for this ratio
yield instead $r_c=1.2$.
Finally, if desired, an estimate of the small $D=6$ contribution
can be included using a rescaled version of the vaccum saturation 
approximation (VSA). This yields
\begin{equation}
\delta R^{L+T}_{OPE;D=6}=-0.0043(1-r_c^2)\left({\frac
{\rho\alpha_s\langle \bar{u}u\rangle^2}{0.000150\ {\rm GeV}^6}}\right)
\end{equation}
where $\rho$ parametrizes VSA breaking. The value in the denominator
of the last factor reflects Ioffe's assessment of $\rho$~\cite{ioffe}.

\end{document}